\documentclass[aps,epsf,twocolumn,showpacs]{revtex4-1}

\usepackage{CJK,amsmath,amssymb,graphics,epsfig,epstopdf,color,verbatim,ulem,braket,tabularx,float}
\usepackage[colorlinks,linkcolor=blue,citecolor=blue,urlcolor=blue]{hyperref}

\begin{document}

\title{Dynamical properties of the Haldane chain with bond disorder}

\author{Jing-Kai Fang}
\email[These authors contributed equally to this work.]{}

\author{Jun-Han Huang}
\email[These authors contributed equally to this work.]{}
	
\author{Han-Qing Wu}
\email{wuhanq3@mail.sysu.edu.cn}

\author{Dao-Xin Yao}
\email{yaodaox@mail.sysu.edu.cn}

\affiliation{Center for Neutron Science and Technology, School of Physics, Sun Yat-sen University, Guangzhou, 510275, China}
\affiliation{State Key Laboratory of Optoelectronic Materials and Technologies, School of Physics, Sun Yat-sen University, Guangzhou, 510275, China}

\begin{abstract}
  By using Lanczos exact diagonalization and quantum Monte Carlo combined with stochastic analytic continuation, we study the dynamical properties of the $S=1$ antiferromagnetic Heisenberg chain with different strengths of bond disorder. In the weak disorder region, we find weakly coupled bonds which can induce additional low-energy excitation below the one-magnon mode. As the disorder increases, the average Haldane gap closes at $\delta_{\Delta}\sim 0.5$ with more and more low-energy excitations coming out. After the critical disorder strength $\delta_c\sim 1$, the system reaches a random-singlet phase with prominent sharp peak at $\omega=0$ and broad continuum at $\omega>0$ of the dynamic spin structure factor. In addition, we analyze the distribution of random spin domains and numerically find three kinds of domains hosting effective spin-1/2 quanta or spin-1 sites in between. These ``spins" can form the weakly coupled long-range singlets due to quantum fluctuation which contribute to the sharp peak at $\omega=0$.
\end{abstract}


\date{\today}
\maketitle

\section{INTRODUCTION}
\label{INTRODUCTION}

The study on the disorder effects of quantum magnetic systems has attracted more and more attentions. From the theoretical point of view, extrinsic disorders can induce spin glass~\cite{SpinGlass}, random-singlet (RS) phase \cite{Fisher1994Random}, Griffths phase ~\cite{Griffiths1}, many-body localization~\cite{MBLAP, MBLRMP}, and new quantum criticality~\cite{Entanglement}. From the experimental point of view, extrinsic disorders make it hard to extract the intrinsic low-energy feature of some quantum phases, like quantum spin liquid phase~\cite{LiuLu2018, HQWu2019, Hikaru2019, Kawamura_2019, MaZhen2020, Muwei2021}. Up to now, it is still a challenging task to numerically study the disorder effect especially on higher-spin systems.

The RS phase which combines randomly coupled two-spin singlets with arbitrary distance has been studied on spin-$1/2$ random antiferromagnetic Heisenberg chain. In this case, the disorder is a relevant perturbation under the renormalization group (RG) transformation which can drive the system into an infinite-randomness fixed point (IRFP)\cite{Infinite2000,Fisher1994Random,Fixpoint}, i.e., any amount of disorder is enough to drive the system into a RS phase\cite{Fisher1994Random,Kawamura_2019}. The strong disorder renormalization group (SDRG) method was introduced to solve the $S=1/2$ random antiferromagnetic Heisenberg chain by Ma, Dasgupta, Hu and Fisher\cite{Ma1979Random,Fisher1994Random,Strong}. This method continues to find two spins with the strongest exchange coupling and decimates the spin pairs from the system, leaving a new effective interaction instead. This renormalization process finally leads the system to its ground state. The SDRG method has been found very successful in the strong disorder case.

\begin{figure}[ht]
	\includegraphics[width=0.45\textwidth]{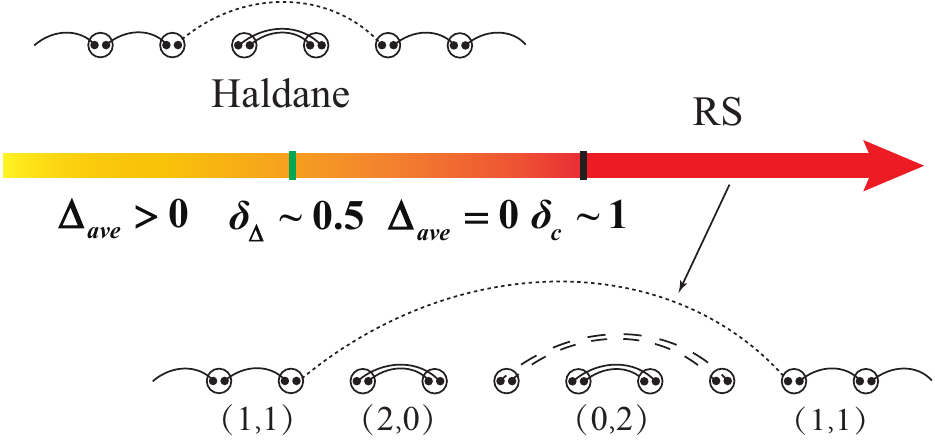}
	\caption{Phase diagram of the random Haldane chain. The randomly averaged gap $\Delta_{ave}$ can survive up to the strength of disorder $\delta\sim 0.5$. $\delta_{c}\sim 1$ is a Haldane-RS critical point after which the system enters the RS phase. For illustration, we plot two skeleton diagrams showing the random Haldane phase and RS phase by following Ref.~\onlinecite{Entanglement}. In the random Haldane phase, the valence-bond-solid ground state breaks up with more and more singlet domains. However, the string order maintains. In the RS phase, (2,0) and (0,2) domains which are singlet dimers $\frac{1}{\sqrt{3}}(\ket{-1,1}-\ket{0,0}+\ket{1,-1})$ on even and odd bonds are significant, and there would be ``isolated" $S=1$ spins in between these two domains. These isolated spins would form long-range singlets (dashed lines) due to quantum fluctuations. At the end of (1,1) domain, there is an ``unpair" $S=1/2$ part, especially at the critical point $\delta_c\sim1$. These $S=1/2$ parts will also form long-range singlets $\frac{1}{\sqrt{2}}(\ket{\uparrow\downarrow}-\ket{\downarrow\uparrow})$ in the renomalization group sense (dotted lines).}
	\label{Fig1}
\end{figure}

Unlike the spin-$1/2$ case, the spin-$1$ antiferromagnetic Heisenberg chain or Haldane chain \cite{Gapped,Haldane1983Nonlinear,HaldaneContinuum} has a finite gap and hidden string order in the pure or clean limit which makes it robust to weak disorder. As shown in Fig. \ref{Fig1}, with the increase of bond disorder, the randomly averaged Haldane gap vanishes and the system enters a quantum Griffiths region \cite{Griffithseffect} with the surviving string order. 
However, if the disorder is strong enough, the system will eventually approach to a Haldane-RS critical point $\delta_{c}$ and then changes into a RS phase\cite{randomsingletphase,spin0.5one,spin0.5twdo}.

\begin{figure*}[ht]
	\centering
	\includegraphics[width=0.8\textwidth]{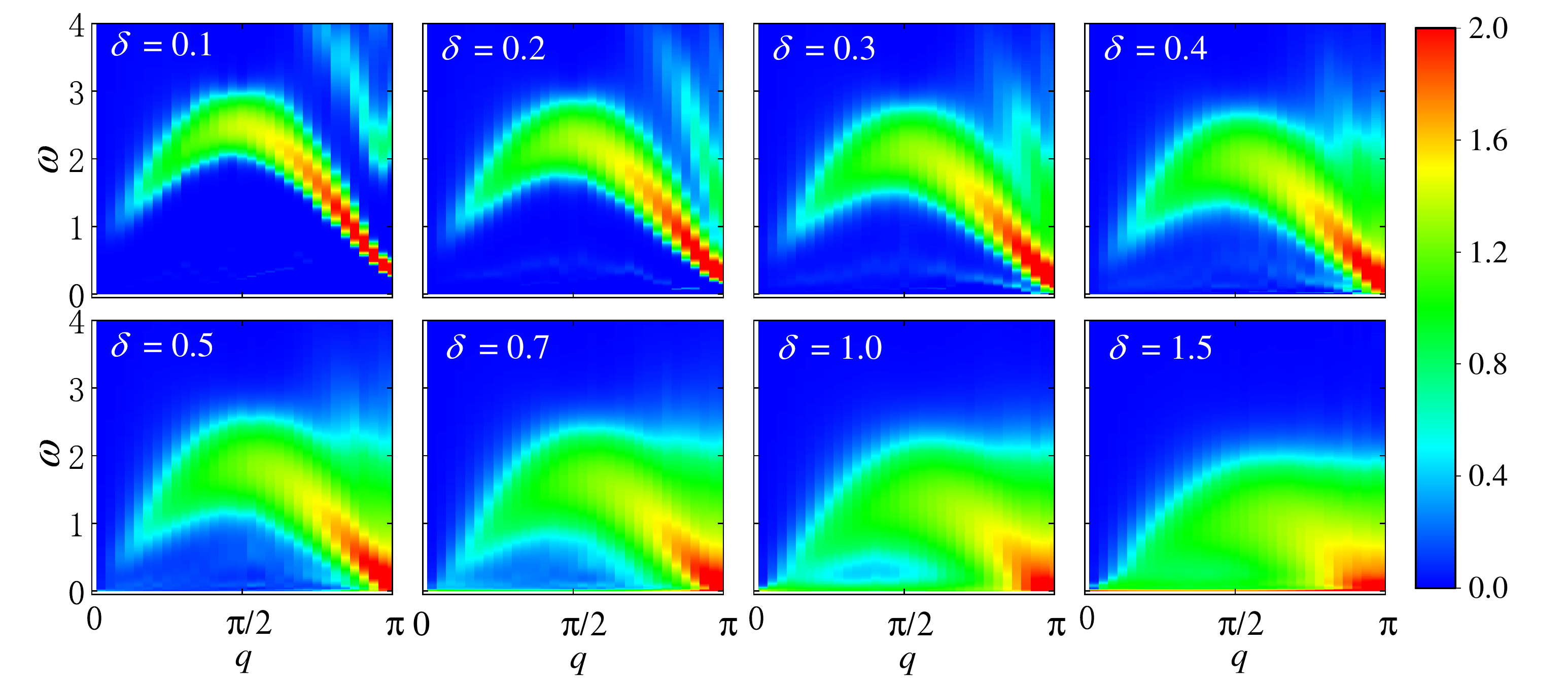}
	\caption{Dynamic spin structure factor $S(q,\omega)$ of the random Haldane chain at different disorder strength. These results are obtained by quantum Monte Carlo combined with stochastic analytic continuation. The system size we used is $L = 64$, and the inverse temperature is $\beta = 128$. For each strength of disorder, we use 5000 different random configurations to get the average.}
	\label{Fig2}
\end{figure*}
 
The presence of disorder makes the numerical calculation extremely strenuous and difficult. The ground-state properties of the random $S=1$ chain has been studied by the extensions of original Ma-Dasgupta rule \cite{Percolation1997,phases,Analytical,Griffithseffect,Analytical,sdrg,Strongly,PhaseDiagram,Quantum,Entanglement}, exact diagonalization (ED) \cite{NISHIYAMA199835}, density matrix renormalization group (DMRG) \cite{HidaDensity,Lajk2005Disorder}, tree Tensor network strong-disorder renormalization group (tSDRG) \cite{Tensor,Self} and quantum Monte Carlo (QMC) \cite{BergkvistGround,Japan}. Despite the complexity of different methods, some ground state properties are very clear. For instance, the average bulk spin correlations \cite{Equal1994} $\overline{C}(r)$ vanish in gapless Haldane phase while average end-to-end correlations $\overline{C_{l}}(L)$ on open chain have a finite limiting value. And both of them decay in power law at the critical point and in the RS phase, i.e. $\overline{C}(r)\sim r^{-\eta}$ and $\overline{C_{l}}(L)\sim L^{-\eta_{l}}$. Even though there have been no theoretical prediction of these exponents at critical point until now, some numerical results have been given in previous works \cite{Lajk2005Disorder,Tensor}. The thermodynamical quantities, including the local magnetic susceptibility $\chi$ and the specific heat $c$, are given by \cite{StrongRgl2005},
 \begin{equation}
 	\chi(T)\sim T^{-1+1/z}, \quad c(T)\sim T^{1/z},
 	\label{eq1}
 \end{equation}
where $z$ is the disorder-induced dynamical exponent \cite{Igl2001Griffiths}.
The previous study of dynamical quantities in the $S=1/2$ Heisenberg chain \cite{YuDynamical,Dynamic0.5,Dynamics0.5three} shows that the presence of disorder can change the $S(q,\omega)$ dramatically. It brings a $\delta$ peak at $\omega \to 0$ in the dynamic spin structure factor $S(q,\omega)$, which is absent in a clean chain. 

Previous numerical studies of random spin-1 chains mainly focus on the ground state properties. In this paper, we investigate the dynamical properties of the $S=1$ random antiferromagnetic Heisenberg chain. We identify that a prominent zero-energy peak also arises at very low frequency in the $S = 1$ chain when the disorder is strong enough even in Haldane phase. Meanwhile, we also present the dynamic spin structure factor $S(q,\omega)$ at week disorder. And there is a special low-energy excitation in the presence of disorder. It is pushed to zero energy and merges with the main broad excitation spectrum when we enhance the strength of bond disorder. The rest of the paper is as follows. In Sec. \ref{MM}, we introduce the model and numerical methods we mainly used. In Sec. \ref{NR}, we show the results of the $S(q,\omega)$ at different strength of disorder. And we try to explain the behavior of the dynamical spectrum by analyze the statistical distributions of spin domains and correlation configurations. In Sec. \ref{Sec:CONCLUSION}, the summary of the paper is given.

\section{MODEL AND METHODS}
\label{MM}

We study the $S=1$ random antiferromagnetic Heisenberg chain (Haldane chain) defined by the Hamiltonian	
\begin{equation}
	\begin{aligned}
	H=\sum_{i}^L J_{i} \ \mathbf{S}_{i}\cdot\mathbf{S}_{i+1},
	\end{aligned}
\end{equation}
where $\mathbf{S}_{i}$ denotes the $S=1$ spin operator on each site $i$ and $J_{i}$ is the random nearest-neighbor exchange coupling. In this paper, we mainly use the power-law distribution of the random exchange couplings to simulate the bond disorder effect,
\begin{equation}
	P_{\delta } (J)=\delta^{-1} J^{-1+1/\delta } \quad \text{for} \ \ 0< J\leq 1
	\label{two}
\end{equation}
where $\delta$ represents the strength of disorder.

The numerical methods we mainly used in this paper are Lanczos exact diagonalization (ED) and quantum Monte Carlo (QMC) combined with stochastic analytic continuation (SAC) \cite{Sandvik2010Computational,Sandvik2002Classical,HeneliusMonte,Alav,Sandvik1999Stochastic,Dorneich2002Accessing}. In ED, we directly calculate the dynamic spin structure factor $S(q,\omega)$ using the Lanczos iterations at zero temperature. However, in quantum Monte Carlo simulation, we cannot directly do that. Instead, we can obtain $S(q,\omega)$ from the imaginary-time correlation function $G_{q}(\tau)$ through the stochastic analytic continuation at a finite but very low temperature. The finite-temperature dynamic spin structure factor is defined in the basis of eigenstates $\ket{n}$ and eigenvalues $E_{n}$ of the Hamiltonian as follows:
\begin{equation}
\begin{split}
	\begin{aligned}
	S^{zz}(q,\omega)= \frac{1}{\mathcal{Z}(\beta)}\sum_{m}\sum _{n}e^{-\beta E_{m}}|\braket{n|S_{q}^{z}|m}|^{2} \times \\\delta [\omega-(E_{n}-E_{m})],
	\end{aligned}
	\end{split}
\end{equation}
where $\mathcal{Z}(\beta)=$ Tr$\{e^{-\beta \mathcal{H}}\}$ is the partition function with inverse temperature $\beta=1/k_{B}T$, and the spin operator $S_{q}^{z}$ is the Fourier transform of the real-space spin operator
\begin{equation}
    \begin{aligned}
    S_{q}^{z}=\dfrac{1}{\sqrt{L}} \sum _{l}e^{-iql}S_{l}^{z}.
    \end{aligned}
\end{equation}
We use the stochastic series expansion QMC to calculate the imaginary-time correlation function
\begin{equation}
	\begin{aligned}
	G_{q}(\tau)=\braket{S_{-q}^{z}(\tau)S_{q}^{z}(0)},
	\end{aligned}
\end{equation}
in which the operator $S_{-q}^{z}(\tau)=e^{\tau \mathit{H}}S_{-q}^{z}(0)e^{-\tau \mathit{H}}$. The relationship between the dynamic spin structure factor and imaginary-time correlation function is given by
 \begin{equation}
 \begin{aligned}
 G_{q}(\tau)=\int _{-\infty}^{\infty}d\omega S(q,\omega)e^{-\tau \omega}.
 \label{eq8}
 \end{aligned}
 \end{equation}	
 Then SAC approach \cite{sacSandvik1998Stochastic,sacSandvikConstrained,sacSandvikHigh,sacshaohui,sacGROSS} is employed in inversing Eq. (\ref{eq8}). It samples the spectrum with a probability distribution like a Boltzmann distribution and fits them to the imaginary-time data. A likelihood function
 \begin{equation}
   \begin{aligned}
   P(S)\propto\exp\left(-\frac{\chi^2}{2\Theta}\right).
  \end{aligned}
 \end{equation}
 is introduced to describe the fitting quality, where $\chi^{2}/2$ represents the energy of a system at a fictitious temperature $\Theta$.

\begin{figure}[t]
	\includegraphics[width=0.42\textwidth]{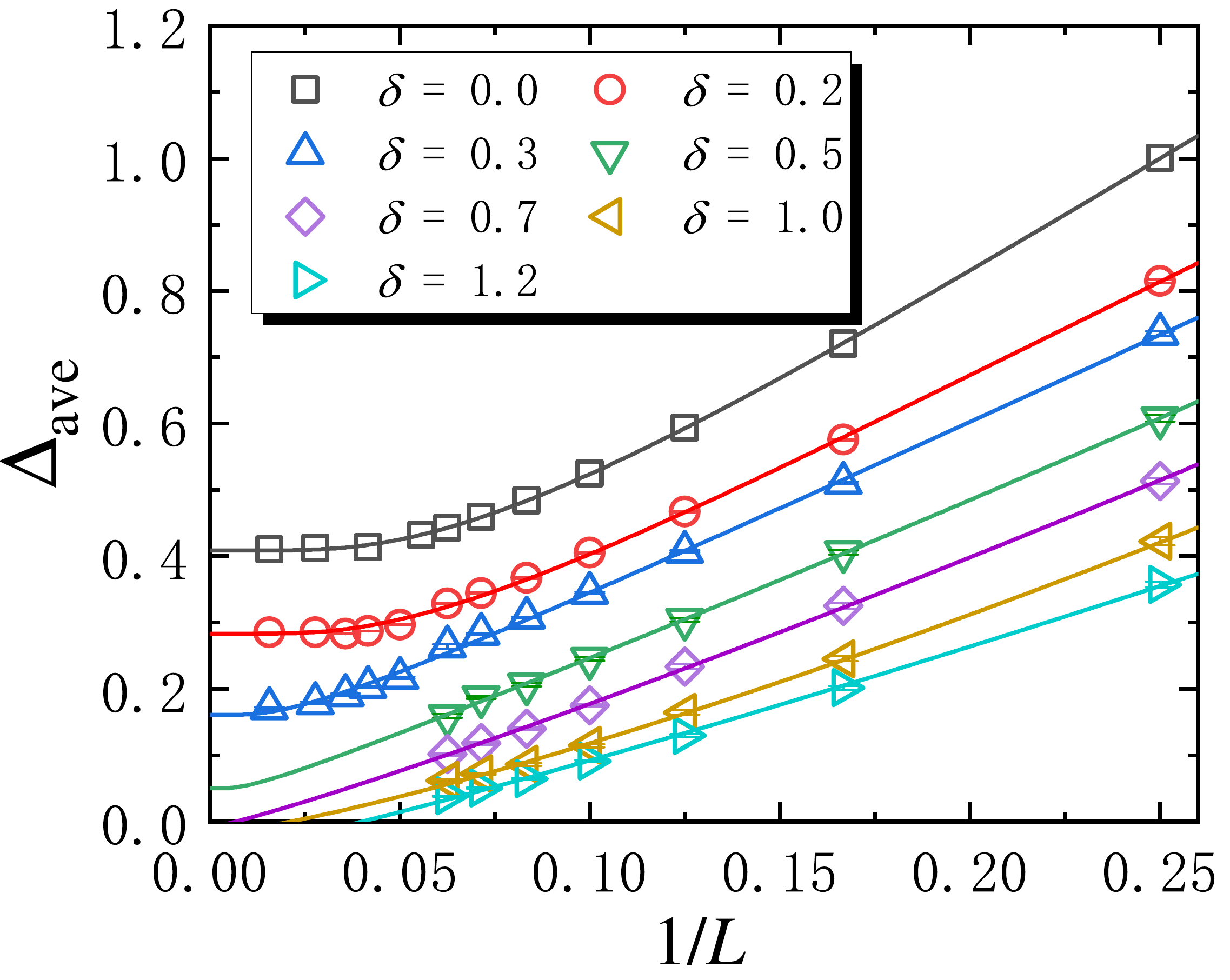}
	\caption{The average triplet gaps with different strengths of disorder as functions of inverse system sizes. For $L\le16$, we use ED to get the triplet gaps. While for other larger system sizes with small $\delta$, we use quantum Monte Carlo data and Eq.~(\ref{eq10}) to fit the gaps. The triplet gap vanishes at $\delta \sim 0.6$. The fitting function is $\Delta(L)=a+e^{-L/\xi}(b/L+c/L^2)$~\cite{GapFitting}. $\xi$ is the correlation length~\cite{CorrLength}.}
	\label{Fig3}
\end{figure}

\section{NUMERICAL RESULTS}
\label{NR}
In this paper, we mainly use $L = 64$ chain with periodic boundary condition and inverse temperature $\beta = 128$ to calculate the dynamic spin structure factor $S(q,\omega)$ in quantum Monte Carlo simulation. For each strength of disorder, 5000 different random bond disorder configurations are employed to get the average values.

We extract the dynamic spin structure factor $S(q,\omega)$ of the $S = 1$ random Haldane chain as shown in Fig.~\ref{Fig2}. In the weak disorder, the single-magnon branch with the minimum gap located at $q=\pi$ is clearly found. With increasing the disorder strength, the average Haldane gap decreases and eventually vanishes at $\delta_{\Delta}\sim 0.5$. To show the closing of the average Haldane gap more clearly, we plot the average Haldane (triplet) gaps with different disorder intensity and various system sizes in Fig.~\ref{Fig3}. For $L\le 16$, we use ED method to get the average triplet gap $\Delta_T=E_1(S=1)-E_0(S=0)$. 
\begin{figure}[t]
    \centering
	\includegraphics[width=0.45\textwidth]{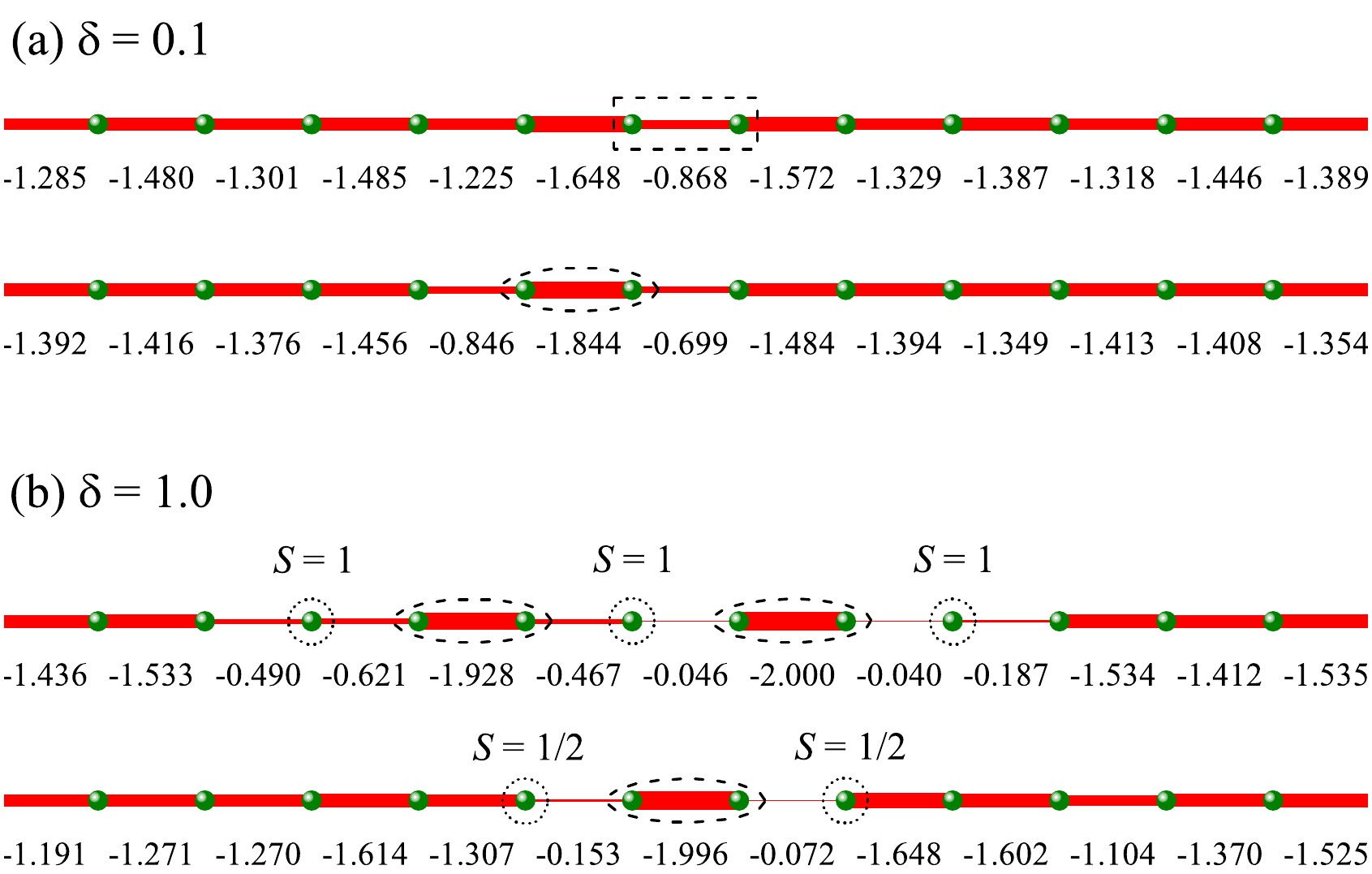}
	\caption{The nearest-neighbor spin correlations (only show some segments of $L=64$) of two typical bond disorder configurations at (a) $\delta=0.1$ and (b) $\delta=1.0$. The dashed rectangle box marks the bond with weak spin correlation. The dashed oval box marks two $S=1$ spins forming a singlet $\frac{1}{\sqrt{3}}(\ket{-1,1}-\ket{0,0}+\ket{1,-1})$. If this singlet forms in a even or odd bond, we call it (2,0) or (0,2) domain. Dotted cycles represent the isolated $S=1$ spins or unpaired $S=1/2$ quanta.}
	\label{Fig4}
\end{figure}
\begin{figure*}[t]
		\centering
	\includegraphics[width=1.01\textwidth]{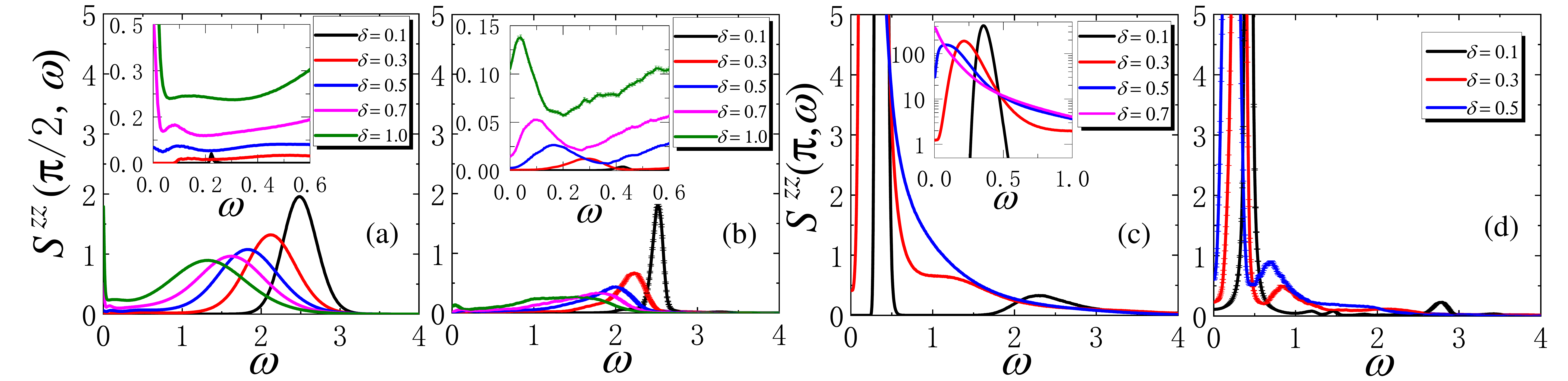}
	\caption{Dynamic spin structure factor $S^{zz}(q,\omega)$ of the random Haldane chain at $q = \pi/2$ and $\pi$. Panels (a) and (c) are calculated by SSE-SAC with $L = 64$ and $\beta = 128$, while panels (b) and (d) show the ED results for $L = 12$ with Lorentz broadening factor $\eta=0.05 J$. Here, we show the results for five different disorder strengths, including $\delta = 0.1$, $\delta = 0.3$, $\delta = 0.5$, $\delta = 0.7$ and $\delta = 1.0$. The insets show more details of the low-energy excitations. In order to see the position of low-energy peak, we use semi-log plot in the inset of (c).}
	\label{Fig5}
\end{figure*}
Here, we want to mention that the lowest eigenvalue is always a singlet for the nonfrustrated chain with even number of sites, which means $\Delta_T$ is always positive. For $L>16$ we use the imaginary-time spin correlation function at transfer momentum point $q=\pi$ to fit the triplet gap using the following equation,
\begin{equation}
G^{zz}_{q=\pi}(\tau) \approx a_0 e^{-\Delta\tau}.
\label{eq10}
\end{equation}

Since this approximation only works well in small $\delta$, we show the fitting gaps for $\delta=0.0, 0.2,0.3$. From Fig.~\ref{Fig3}, we find that the closing point of average triplet gap is $0.5<\delta_\Delta<0.7$ which is larger than the one got from $S(q,\omega)$ due to the limiting size of ED method. Therefore, in the small bond disorder region $\delta<0.5$, there is still a finite average triplet gap. The results are consistent with previous numerical results obtained by DMRG \cite{Lajk2005Disorder}. But the single-magnon branch decays and becomes broad. When looking closer at the region below the single-magnon branch, we find a new weak spectrum in the upper four figures of Fig.~\ref{Fig2}. What is the origin of this spectrum? To understand that we can go back to the spin-1 AKLT state which can adiabatically connected to Haldane phase. In that state, $S=1$ can decomposed into symmetrized product of two $S=1/2$ quanta. And two effective spin-1/2 placed on adjacent sites form singlets in the AKLT state. Therefore, the lowest excitation can be seen as a propagating bond triplet. Under weak bond disorder, the exchange interactions randomly distribute, some nearest-neighbor bonds with weak interactions contribute to the lower-energy nearly flat excitation. In Fig.~\ref{Fig4}(a), we show the nearest-neighbor spin correlations under a typical bond randomness configuration with $\delta=0.1$. We can find that some of nearest-neighbor spin correlations departure from mean value due to the weaker exchange interactions. And these weakly singlets formed by two nearby $S=1/2$ quanta are more easily excited to be triplets which contribute to the new low-energy excitation below the single-magnon mode. The random distribution of weak bonds make the low-energy spectrum dispersive.

\begin{figure}[h]
	\includegraphics[width=0.49\textwidth]{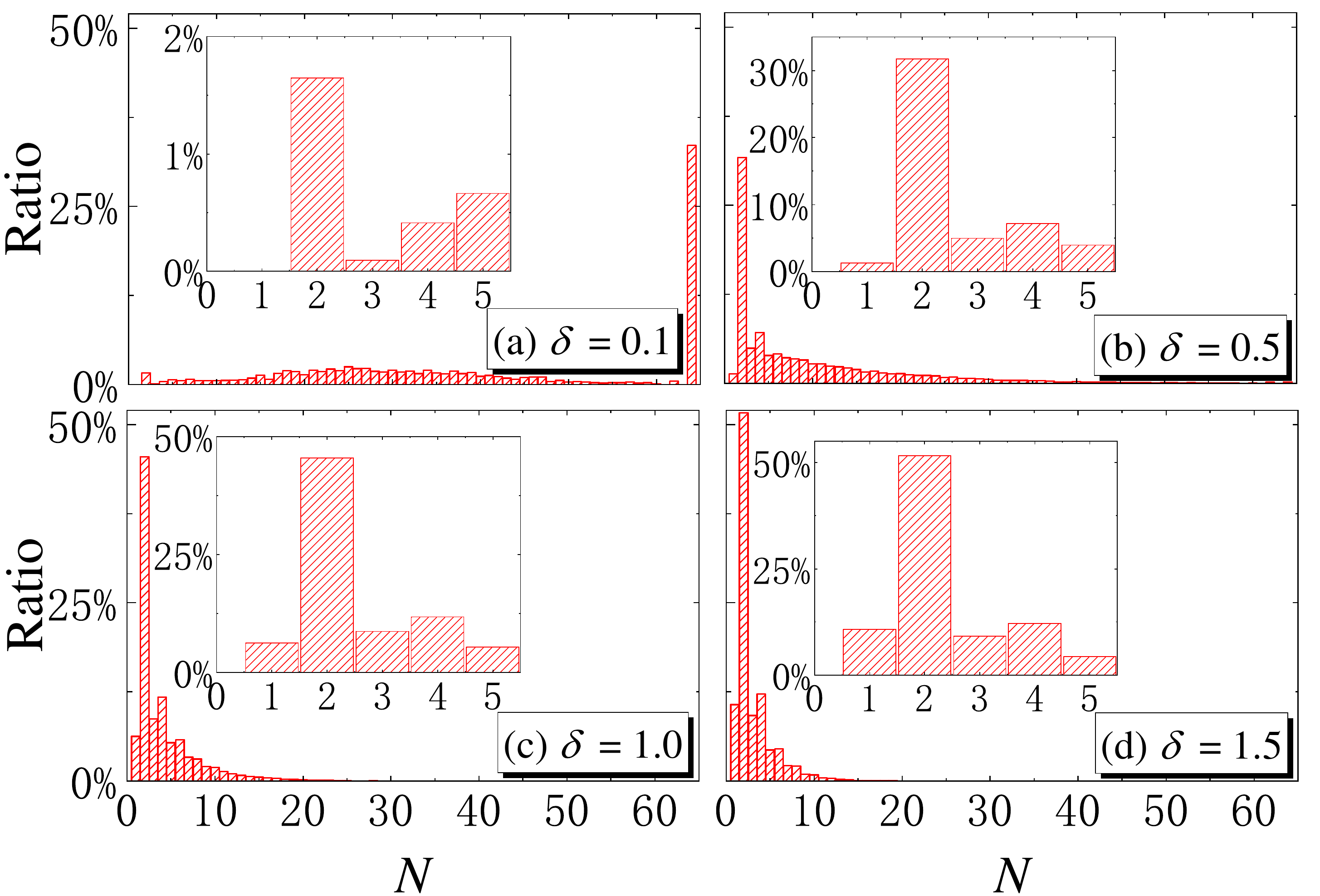}
	\caption{Histograms of spin domains at different strength of disorder $\delta = 0.1$, $0.5$, $1.0$ and $1.5$. The results are calculated using $L = 64$ and 2000 random configurations at each disorder intensity. The insets show the ratios of spin domains with small sites. }
	\label{Fig6}
\end{figure}

To see more clearly about the weak-bond excitations, we show some 1D cuts of the dynamic spin structure factor at transfer momentum $q=\pi/2$. In Fig. \ref{Fig5}(a) and Fig. \ref{Fig5}(b), we extract the dynamic spin structure factor $S(\pi/2, \omega)$ with different strength of disorder and from two kinds of methods, i.e. SSE-SAC and ED. Unlike the random $S=1/2$ chains, the zero-energy sharp peak absents at weak disorder. The insets show the detail of low-energy excitation, which is attributed to weak singlets on weak bonds. As the disorder strength $\delta$ increases, the low-energy peak becomes broad and shifts to zero energy. Meanwhile, its spectral weight becomes stronger and stronger. Because of the enhancement of randomness, the number of weaker nearest-neighbor interactions grows and the strength of these interactions becomes weaker and weaker. Thus, these weak bonds are more easy to be excited, and their spectral weights increase at the same time. In Fig. \ref{Fig5}(c) and Fig. \ref{Fig5}(d), we also show the 1D cuts of the dynamic spin structure factor $S(\pi,\omega)$ at transfer momentum $q=\pi$ with three different strengths of disorder. We can observe a broad continuum at higher energy. In the weak strength of disorder $\delta = 0.1$, the excitation spectrum consists a single-magnon branch and notably weak multimagnon continuum at higher energy \cite{multimagnon, multimagnon2, multimagnon3, Spectral, Finite, Dynamical}, which is similar to a clean Haldane chain. With the increase of disorder strength [see $\delta = 0.3$ and $0.5$ in Figs. \ref{Fig5} (c) and (d)], the multimagnon continuum gradually approach the prominent single-magnon peak and merges into a broad continuum at $q=\pi$ finally. Thus, we can clearly see that the spectral weight is pushed to $\omega \approx 0$ due to increasing numbers of weaker random bonds with the increase of disorder strength.

\begin{figure}[t]
	\includegraphics[width=0.47\textwidth]{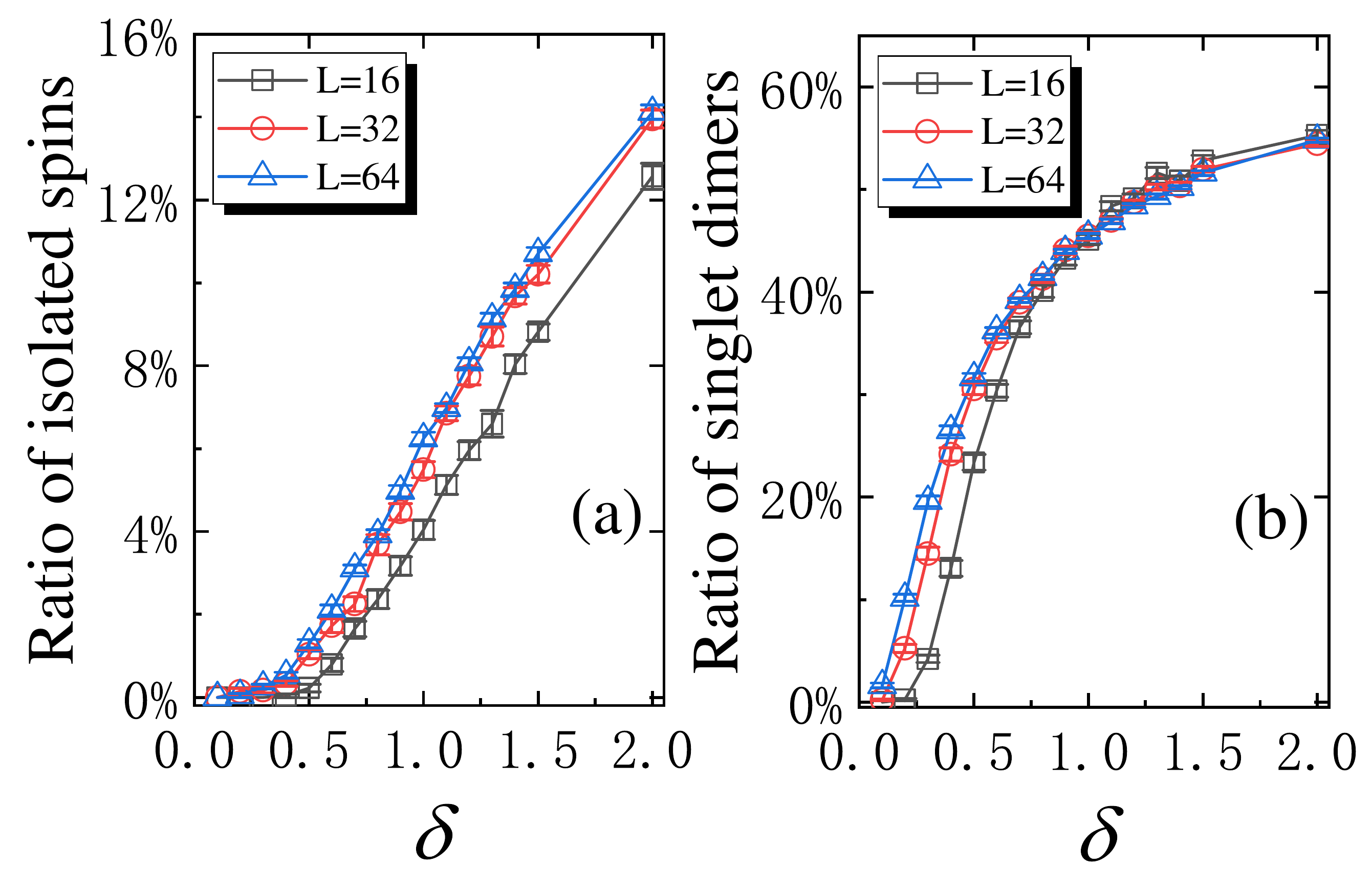}
	\caption{The ratios of (a) isolated $S=1$ spin and (b) singlet dimers at different disorder intensities. We have used three system sizes $L = 16$, $32$, and $64$ to do the calculations.}
	\label{Fig7}
\end{figure}

When further increasing the disorder strength, the average triplet gap $q=\pi$ closes, and the system enters the gapless Griffths region with the surviving hidden string order. For the dynamic spin structure factor, the flat low-energy excitation below the single-magnon mode touches the wall at $\omega = 0$, and the zero-energy peak arises and its intensity increases quickly, as can be seen in Fig.~\ref{Fig2} and the insets of Fig.~\ref{Fig5}. More and more spectrum weight shifts to the low-energy part, and the single-magnon mode becomes even more broad.

At the Haldane-RS critical point $\delta = 1$, the excitation spectrum becomes similar to the strong disorder case $(\delta = 1.5)$ and $S=1/2$ case. The whole spectrum is a broad continuum. And the zero-energy sharp peak becomes dominant effect on low-energy physics. To give a better understanding of the excitation at this critical point, we show the nearest-neighbor spin correlations under some typical random configurations in Fig. \ref{Fig4}(b). We can see the ``isolated" $S=1$ spins between (2,0) and (0,2) singlet domains. In addition, we can still identify some unpaired spin-1/2 quanta at the ends of (1,1) domain \cite{Entanglement}. The random distribution of these three kinds of spin singlet domains gives rise to the isolated $S=1$ spin and unpaired $S=1/2$. These $S=1$ spins and $S=1/2$ quanta can form long-range singlets due to the quantum fluctuation with very low energy, see Fig.~\ref{Fig1}. Therefore, it costs very low energy to break up these singlets. These excitations contribute to the spectrum near zero-energy. In the RS phase with larger $\delta$, the (2,0) and (0,2) domains accompanied by isolated $S=1$ spins are typical characteristics of $S=1$ RS phase.

 In order to find how the distribution of spin domains changes with the disorder strength, we show the histogram of all the spin domains at different $\delta$. The nearest-neighbor spin correlation functions $\braket{\mathbf{S}_{i}\cdot \mathbf{S}_{i+1}}$ [with the range $(-2.0,0.0)$] are calculated to find the spin domains. For each disorder strength with $L=64$, we use 2000 random configurations to calculate the average values. We set a cutoff value $C = -1.0$ (half of the range) as the criterion of whether two neighboring spins are in the same domain or not. If the spin correlation of two neighboring spins is less than $C$, then these two spins are considered to be in the same domain. The histograms of spin domains with $N$ sites are shown in Fig.~\ref{Fig6}. In the weak disorder region $\delta=0.1$, $N=L$ is found to be the most possible spin domains, which means the system is close to the pure Haldane chain. However, we still find some effective two $S=1$ spin singlet dimers sandwiched between two weakly coupled bonds under some random configurations, as can be seen in Fig.~\ref{Fig4}(a). In addition, two unpaired $S=1/2$ quanta at the ends of two long chains can form effective $S=1/2$ singlets mediated by quantum fluctuation, or they can be ``resonant" with $S=1$ singlet dimers to form weakly coupled dimers. They contribute to the low-energy excitation below the one-magnon mode. As the disorder strength increases, the whole chain is ``divided into" more smaller spin domains, and the singlet dimers become more prominent. Meanwhile, the ratio of isolated $S=1$ spins stuck in the middle of two singlet dimers also increases. In Fig.~\ref{Fig7}, we show the ratios of isolated spin and singlet dimer with increasing the disorder strength $\delta$. To consider the possible finite-size effect, we also show the results of two smaller system sizes $L=16$ and $L=32$. From the results, we can conclude that $L=64$ is large enough for the thermodynamic limit. Around $\delta=0.5$, the ratio of isolated spins increases significantly, where the average energy gap is closing. For the singlet dimers, a weak $\delta$ could induces some finite distributions. When $L$ is larger, the singlet dimers appear more easily.  After the critical point $\delta\sim 1$, we find the finite-size effect is weak because the system is in the RS phase in which the singlet dimers dominate.

\section{DISCUSSION AND CONCLUSION}
\label{Sec:CONCLUSION}

In summary, we have numerically studied the dynamical properties of the random Haldane chain by calculating the randomly averaged dynamic spin structure factor $S(q,\omega)$. From the previous works, we have a clear picture of ground state properties. In contrast to the random $S = 1/2$ Heisenberg chain, the random Haldane chain does not convert directly to the RS phase in the presence of weak disorder. The system remains in the Haldane phase with a finite average gap. The dynamic spin structure factor is similar to the pure Haldane phase, but with the emerging low-energy excitation below the one-magnon mode. As the disorder strength increases, the average Haldane gap closes and the system enters a gapless Haldane phase with more and more low-energy excitation. When the disorder is strong enough, the system arrives at a critical point $\delta_{c}$, where the effective long-range $S=1$ and $S=1/2$ singlets can be formed by two isolated spins or two ends of Haldane chain segments. In the RS phase, we show that the (2,0) and (0,2) domains are mainly accompanied by the isolated $S=1$ spins and the dynamic spin structure factor becomes broad continuum with prominant zero-energy spectral weight originated from the weakly coupled long-range singlets of isolated spins. Furthermore, we have studied the statistical distributions of spin domains and correlation configurations which can help to understand the theoretical scenarios and the numerical spectra.

Finally, it is a very interesting topic of the higher-spin case. For a sufficiently strong disorder, all the ground states of the random antiferromagnetic Heisenberg spin-$S$ chains become a collection of nearly independent singlets of spin-$S$ pairs with arbitrarily range in real space. The low energy spectrum associated with these singlets is extremely broad with prominent zero-energy spectral weight governed by the weakly-coupled quite-distant singlets. However, the weak disorder case is more complicated. There are two cases we need to address according to Haldane conjecture. One is the integer Heisenberg spin chains like $S = 1$, 2, … Haldane chain. The finite excitation gap in the clean case prevents the formation of random singlet phase immediately when introduced bond disorder. Another is the half-integer Heisenberg spin chains like $S=1/2$, $3/2$, … chain. It is clear that, for $S = 1/2$, the infinitesimal bond disorder can flow to the infinite randomness fix point. But some theoretical studies \cite{spin32Refael,Damle32} show that there may be a sequence of effective spin $S$, $S-1$, ..., 1/2 random singlet phase between strong RS phase (spin $S$) and clean phase. Take $S = 3/2$ as an example, Ref.\cite{spin32Refael} points out that the weak disorder phase is also an RS phase which the ground state is an effective spin-1/2 chain superimposed on a Haldane phase. Some other studies \cite{Disorderinducedphasesinhigherspin,2005Strong,Saguia32} even detected an irrelevant perturbation region at weak disorder. In the higher spin case, SDRG may not fully capture the physics in the intermediate bond disorder \cite{Breakdown}. Therefore, it is still worthy to explore them using unbiased numerical methods in the future.

\begin{acknowledgments}
 We thank Yu-Cheng Lin, Anders W. Sandvik, and Muwei Wu for helpful discussions. J.K.F. acknowledges Yu-Cheng Lin for three months of host as exchange study at National Chengchi University. D.X.Y., J.K.F, and J.H.H are supported by NKRDPC-2017YFA0206203, NKRDPC-2018YFA0306001, NSFC-11974432, GBABRF-2019A1515011337, National Supercomputer Center in Guangzhou and Leading Talent Program of Guangdong Special Projects. H.Q.W. is supported by NSFC-11804401 and the Fundamental Research Funds for the Central Universities, Sun Yat-sen University (Grant No. 2021qntd27).
\end{acknowledgments}

\appendix
\section{Calculation of energy gap}

\begin{figure}[h]
	\centering
	\includegraphics[width=0.47\textwidth]{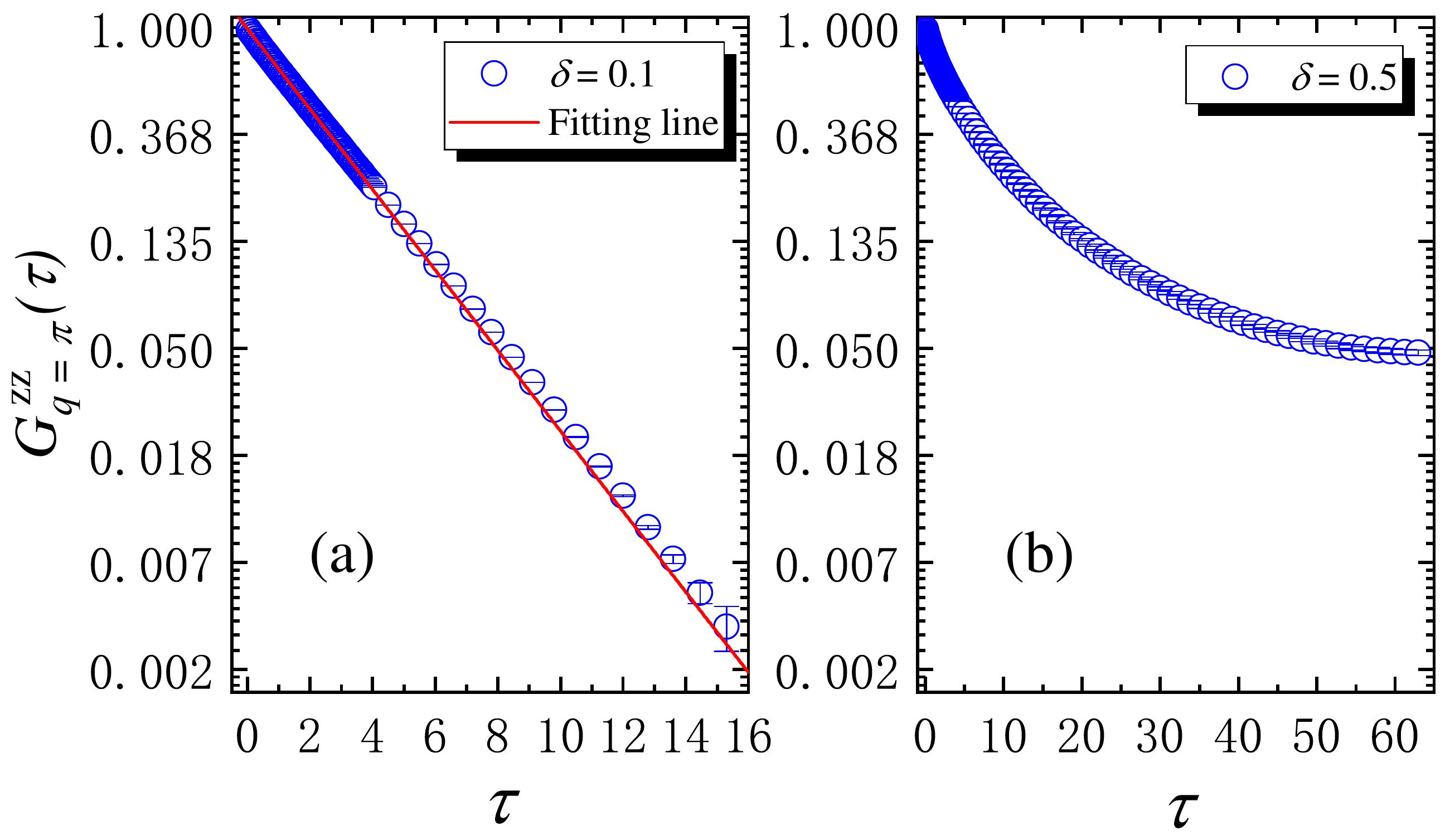}
	\caption{The normalized imaginary-time correlation function $G^{zz}_{q=\pi}(\tau)$ at momentum $q = \pi$ for the $L = 64$, $ \beta =128$ random Heisenberg chain at (a) $\delta = 0.1$ and (b) $\delta = 0.5$. The red straight line is the fitting corresponds to the Eq. (\ref{eq10}) with $a_{0}=0.98900(12)$ and $\Delta \approx 0.37611(24)$. However, the fitting fails for some sufficient strong disorder strengths. }
	\label{Fig8}
\end{figure}

For the weak disorder case, the average energy gap survives under randomness. We can simply obtain the triplet gap by fitting the imaginary-time correlation function using Eq. (\ref{eq10}), where $a_{0}$ is the amplitude of the prominent peak and $\Delta$ is the average energy gap. Here, we show the imaginary-time correlation function at $\delta = 0.1$ (with linear fitting) and $\delta =0.5$ in Fig. \ref{Fig8}. The imaginary-time correlation function can not be fitted well when the disorder becomes stronger, especially for $\delta\sim0.5$ where the energy gap closes. The imaginary-time correlation function becomes nonlinear at semi-log plot which leads to an obstacle for us to obtain an accurate energy gap. Thus, we mainly use ED to calculate the average energy gap in the strong disorder case.

\section{Susceptibility}

\begin{figure}[ht]
	\includegraphics[width=0.40\textwidth]{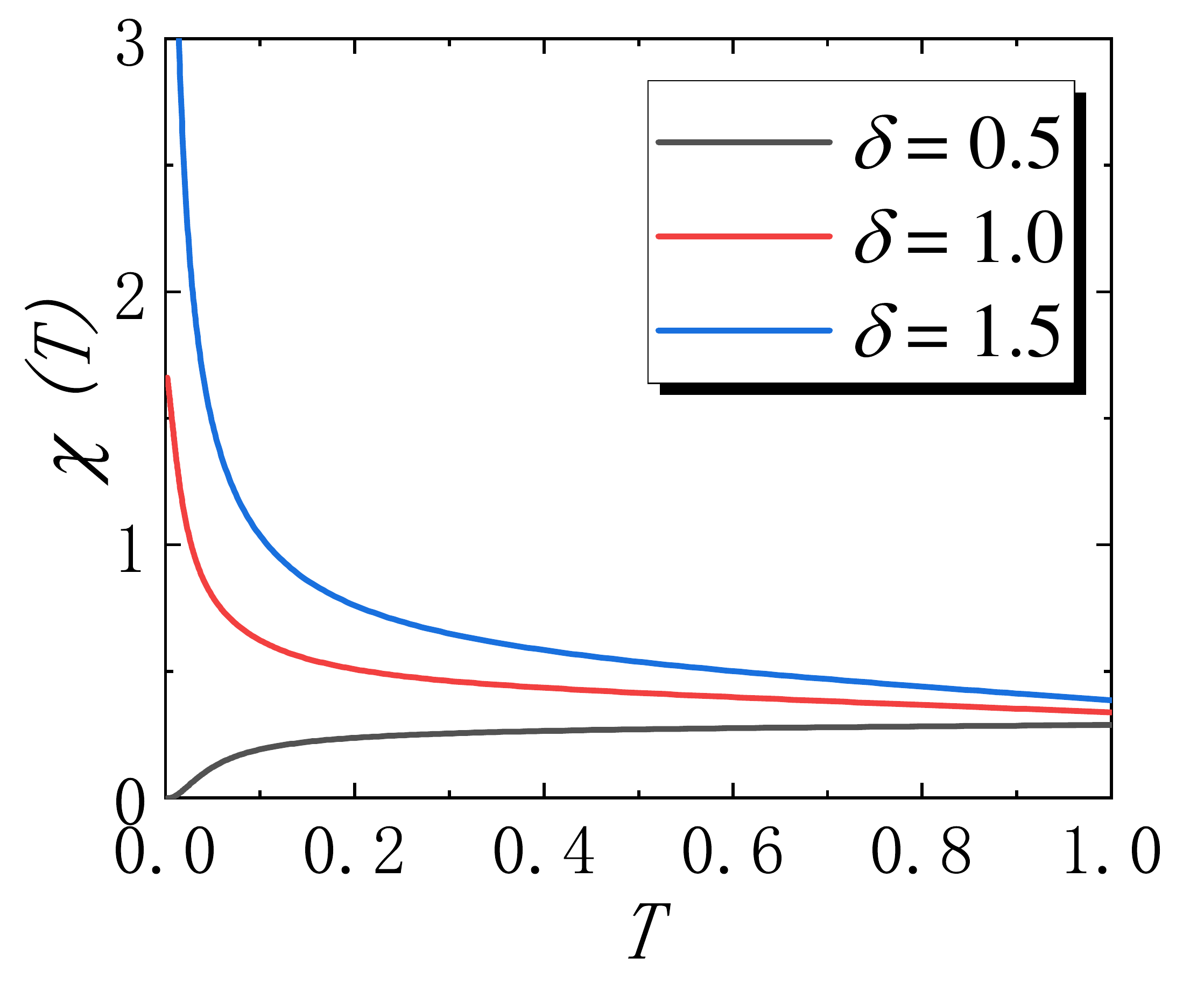}
	\caption{The local magnetic susceptibility $S = 1$ Heisenberg chains with different randomness.}
	\label{Fig9}
\end{figure}

The distinctness of the excitation spectrum between the different strengths of disorder at Haldane phase can also be explained by the separation of the dynamical exponent $z < 1$ and $z > 1$. From the Eq. (\ref{eq1}), the $\chi(T)$ will vanish with $z < 1$ and diverge with $z > 1$ \cite{Lajk2005Disorder}. The local magnetic susceptibility $\chi(T)$ with different strengths of disorder $(\delta = 0.5, \delta = 1, \delta = 1.5)$ is shown in Fig. \ref{Fig9}. At low temperature $T \to 0$, $\chi(0)$ vanishes at $\delta = 0.5$ and diverges at $\delta = 1$ and $1.5$. The gapless Haldane phase can be divided into two regions based on the dynamical exponent $z$. In the region without singularity, $z < 1$, the average energy gap $\Delta_{ave}$ is finite, and vanishes in the singular region, $z>1$.

The divergent behavior of the local magnetic susceptibility at $T \to 0$ is related to the high density of the low-energy excitation spectrum at low frequency $(\omega\to 0)$ \cite{Fisher1994Random}, shown in Fig. \ref{Fig2}. With the vanishing of average energy gap and the divergence of local magnetic susceptibility, the gapless Haldane phase is separated into two different regions. The low energy excitation grows rapidly in the $z >1$ region, similar as the RS case even though they are in the different phases.

\bibliography{reference}

\end{document}